\documentclass[aps,prl,groupedaddress,twocolumn,nopacs,nobibnotes,amsmath,amssymb,letter]{revtex4-1}
\setcitestyle{super}
\usepackage{color}
\usepackage{graphicx}
\usepackage{dcolumn}
\usepackage{bm}
\usepackage{ulem}
\usepackage{etoolbox} 
\newcounter{firstbib} 

\AtBeginEnvironment{thebibliography}{
  
}

\apptocmd{\thebibliography}{
  \setcounter{NAT@ctr}{\value{firstbib}} 
}{}{}  

\AtEndEnvironment{thebibliography}{
  \setcounter{firstbib}{\value{NAT@ctr}}
}

\begin{document}
\title{Ubiquitous formation of bulk Dirac cones and topological surface states from a single orbital manifold in transition-metal dichalcogenides}

\author{M.~S.~Bahramy$^{1,2,*,\dag}$, O.~J.~Clark$^{3,*}$, B.-J.~Yang$^{4,5,6}$, J.~Feng$^{3,7}$, L.~Bawden$^3$, J.~M.~Riley$^{3,8}$, I.~Markovi{\'c}$^{3,9}$, F.~Mazzola$^3$, V.~Sunko$^{3,9}$, D.~Biswas$^3$, S.~P.~Cooil$^{10}$, M.~Jorge$^{10}$, J.~W.~Wells$^{10}$, M.~Leandersson$^{11}$, T.~Balasubramanian$^{11}$, J.~Fujii$^{12}$, I.~Vobornik$^{12}$, J.~E.~Rault$^{13}$, T.~K.~Kim$^{8}$, M.~Hoesch$^8$, K.~Okawa$^{14}$, M.~Asakawa$^{14}$, T.~Sasagawa$^{14}$, T.~Eknapakul$^{15}$, W.~Meevasana$^{15,16}$, P.~D.~C.~King$^{3,\ddag}$}

\affiliation{$^1$ Quantum-Phase Electronics Center and Department of Applied Physics, The University of Tokyo, Tokyo 113-8656, Japan}
\affiliation{$^2$ RIKEN center for Emergent Matter Science (CEMS), Wako 351-0198, Japan}
\affiliation{$^3$ SUPA, School of Physics and Astronomy, University of St. Andrews, St. Andrews, Fife KY16 9SS, United Kingdom}
\affiliation{$^4$ Department of Physics and Astronomy, Seoul National University, Seoul 08826, Korea}
\affiliation{$^5$ Center for Correlated Electron Systems, Institute for Basic Science (IBS), Seoul 08826, Korea}
\affiliation{$^6$ Center for Theoretical Physics (CTP), Seoul National University, Seoul 08826, Korea}
\affiliation{$^{7}$ Suzhou Institue of Nano-Tech and Nanobionics (SINANO), CAS, 398 Ruoshui Road, SEID, SIP, Suzhou, 215123, China}
\affiliation{$^8$ Diamond Light Source, Harwell Campus, Didcot, OX11 0DE, United Kingdom}
\affiliation{$^9$ Max Planck Institute for Chemical Physics of Solids, N{\"o}thnitzer Stra{\ss}e 40, 01187 Dresden, Germany}
\affiliation{$^{10}$ Center for Quantum Spintronics, Department of Physics, Norwegian University of Science and Technology, NO-7491 Trondheim, Norway}
\affiliation{$^{11}$ MAX IV Laboratory, Lund University, P. O. Box 118, 221 00 Lund, Sweden}
\affiliation{$^{12}$ Istituto Officina dei Materiali (IOM)-CNR, Laboratorio TASC, in Area Science Park, S.S.14, Km 163.5, I-34149 Trieste, Italy}
\affiliation{$^{13}$ Synchrotron SOLEIL, CNRS-CEA, L'Orme des Merisiers, Saint-Aubin-BP48, 91192 Gif-sur-Yvette, France}
\affiliation{$^{14}$ Laboratory for Materials and Structures, Tokyo Institute of Technology, Kanagawa 226-8503, Japan}
\affiliation{$^{15}$ School of Physics and Center of Excellence on Advanced Functional Materials, Suranaree University of Technology, Nakhon Ratchasima, 30000, Thailand}
\affiliation{$^{16}$ ThEP, Commission of Higher Education, Bangkok 10400, Thailand}
\affiliation{}
\affiliation{$^{*}$ These authors contributed equally to this work}
\affiliation{$^{\dag}$ To whom correspondence should be addressed: bahramy@ap.t.u-tokyo.ac.jp}
\affiliation{$^{\ddag}$ To whom correspondence should be addressed: philip.king@st-andrews.ac.uk}

\begin{abstract}
{\bf Transition-metal dichalcogenides (TMDs) are renowned for their rich and varied bulk properties, while their single-layer variants have become one of the most prominent examples of two-dimensional materials beyond graphene. Their disparate ground states largely depend on transition metal $d$-electron-derived electronic states, on which the vast majority of attention has been concentrated to date. Here, we focus on the chalcogen-derived states. From density-functional theory calculations together with spin- and angle-resolved photoemission, we find that these generically host a co-existence of type-I and type-II three-dimensional bulk Dirac fermions as well as ladders of topological surface states and surface resonances. We demonstrate how these naturally arise within a single $p$-orbital manifold as a general consequence of a trigonal crystal field, and as such can be expected across a large number of compounds. Already, we demonstrate their existence in six separate TMDs, opening routes to tune, and ultimately exploit, their topological physics.}
\end{abstract}

\date{\today}
            
\maketitle 

\begin{figure}[!h]
\includegraphics[width=\columnwidth]{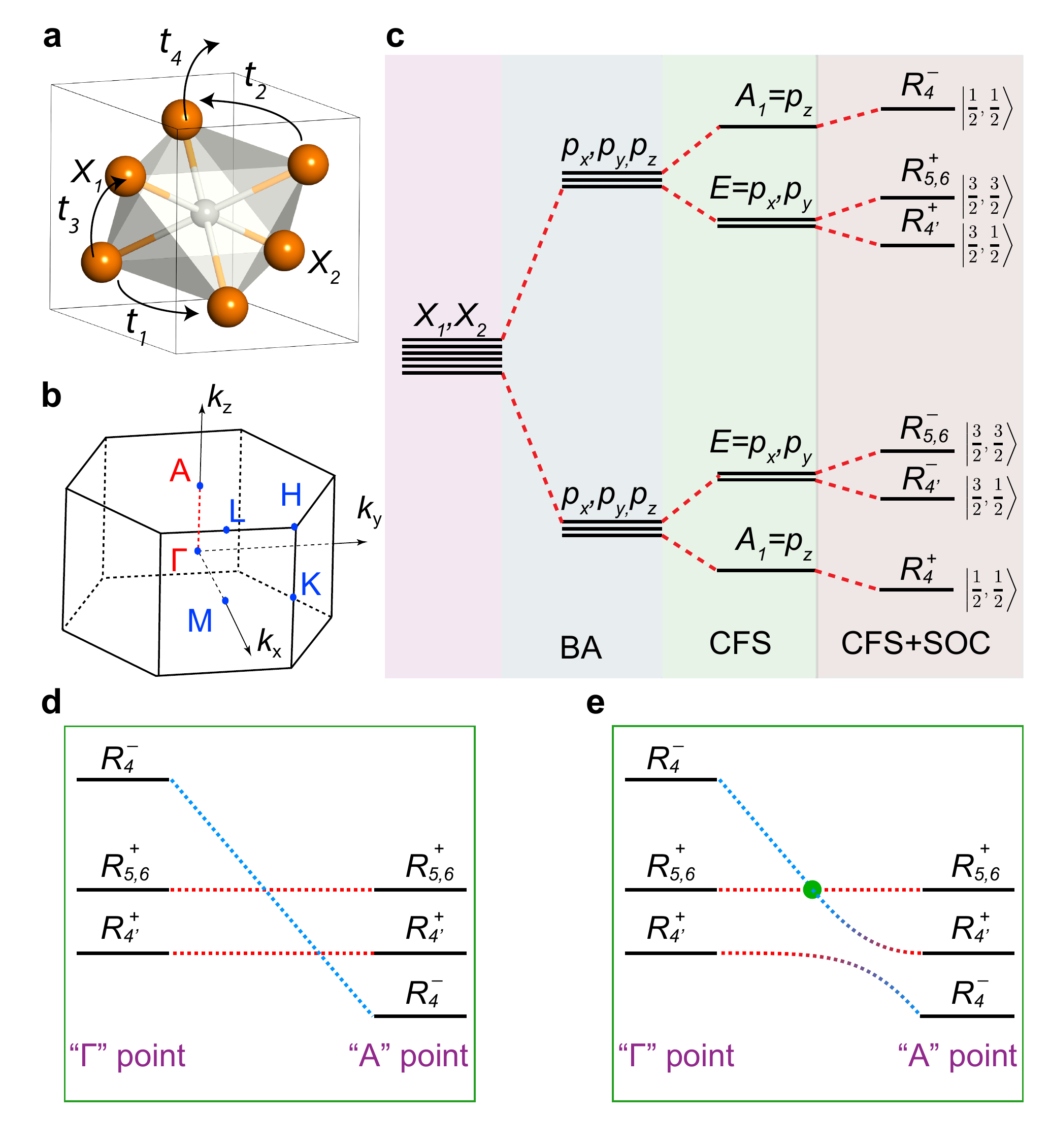}
\caption{ \label{f:overview} {\bf Hierarchy of band inversions arising from $p$ orbitals in a trigonal crystal field.} (a) Crystal structure of the $1$T transition-metal dichalcogenides, with the transition metal at the centre of a trigonally-distorted octahedron of chalcogen atoms ($X_1$ and $X_2$) which each form triangular layers above and below the transition-metal plane. The corresponding Brillouin zone is shown in (b). (c) Schematic illustration of the orbital energy level evolution of $p$-orbitals: Bonding and anti-bonding (BA) combinations form triply-degenerate (neglecting spin) manifolds, which are each split by the trigonal crystal field (CFS) into a doubly-degenerate $E$ level (formed from $p_x$ and $p_y$ orbitals) above (bonding) or below (anti-bonding) a single $A_{1}$ ($p_z$-derived) level. Including spin-orbit coupling (SOC), the $A_{1}$ level transforms into the double representation $R_{4}^\pm$ ($|J\!=\!1/2; |m_J|\!=\!1/2\rangle$) and the $E$ manifold further splits into an upper $R_{5,6}^\mp$ ($|J\!=\! 3/2; |m_J|\!=\!3/2\rangle$) and a lower $R_{4\prime}^\mp$ ($|J\!=\!3/2; |m_J|\!=\!1/2\rangle$) level. The superscript $\pm$ indicates the parity of each level which can be either $+$ (even parity) or $-$ (odd parity) depending on its bonding nature. (d,e) Evolution of these crystal field-derived levels (anti-bonding set) as a function of out-of-plane momentum, showing a crossing of the $A_{1}$ and $E$-derived levels that is naturally expected due to their disparate out-of-plane dispersion. Hybridisation is neglected in (d) but included in (e), showing the resulting formation of a protected crossing and the opening of an inverted band gap with $\mathbb{Z}_2$ topological order at the crossing of the $R_{4}^-$ level with the $R_{5,6}^+$ and $R_{4\prime}^+$ levels, respectively. Hopping paths considered in our tight-binding model shown in Fig.~\ref{f:phase} are indicated schematically in (a).}
\end{figure}

The classification of electronic structures based on their topological properties has opened powerful routes for understanding solid state materials.~\cite{hasan_colloquium:_2010} The now-familiar $\mathbb{Z}_2$ topological insulators are most renowned for their spin-polarised Dirac surface states residing in inverted bulk band gaps.~\cite{hasan_colloquium:_2010} In systems with rotational invariance, a band inversion on the rotation axis can generate protected Dirac cones with a point-like Fermi surface of the bulk electronic structure.~\cite{young_dirac_2012, wang_dirac_2012, wang_3D_2013,borisenko_experimental_2014,liu_discovery_2014,yang_classification_2014,yang_prb_2015} If either inversion or time-reversal symmetry is broken, a bulk Dirac point can split into a pair of spin-polarised Weyl points.~\cite{wan_topological_2011,xu_discovery_2015,yang_weyl_2015,lv_observation_2015,lv_experimental_2015,weng_weyl_2015,borrisenko_T_Weyl} Unlike for elementary particles, Lorentz-violating Weyl fermions can also exist in the solid state, manifested as a tilting of the Weyl cone.  If this tilt is sufficiently large, so-called type-II Weyl points can occur, now formed at the touching of open electron and hole pockets.~\cite{soluyanov_type-ii_2015, Xu_structured, borrisenko_T_Weyl,huang_spectroscopic_2016,Deng_experimental_2016,Tamai,obrien_magnetic_2016,mccormick_minimal_2017}

Realising such phases in solid-state materials not only offers unique environments and opportunities for studying the fundamental properties of fermions, but also holds potential for applications exploiting their exotic surface excitations and bulk electric and thermal transport properties.~\cite{chiral_anomoly,axial,ferreiros_anamalous_2017,Saha_anomalous_2017,mccormick_semiclassical_2017} Consequently, there is an intense current effort focused on identifying compounds which host the requisite band inversions. In many cases, however, this depends sensitively on fine details of a material's electronic or crystal structure.  This is partly because almost all known topologically non-trivial phases are stabilised by inversions between states derived predominantly from different atomic manifolds in two- (or more) component compounds (e.g. Bi and Se $p$ orbitals in Bi$_2$Se$_3$;~\cite{zhang_topological_2009} Bi $p$ and Na $s$ orbitals in Na$_3$Bi;~\cite{wang_dirac_2012} Nb $d$ and P $p$ orbitals in NbP~\cite{belopolski_criteria_2016}). In contrast, here we uncover a simple and remarkably-robust mechanism for realising a hierarchy of band inversions within a single orbital manifold. Across the broad family of 2H- and 1T-structured transition-metal dichalcogenides (TMDs)~\cite{wang_electronics_2012,chhowalla_2013,xu_spin_2014}, we observe and classify how this mediates the formation of strongly-tilted type-I and type-II bulk Dirac cones as well as ladders of topological surface states (TSSs) and topological surface resonances. 

\

\noindent{\bf Band inversions from a single orbital manifold}\\
Figure~\ref{f:overview} details the general principle underlying our findings. As a minimal model, we consider a 2-site system with space group $C_{3v}$, with $3\times2$ $p$-orbitals per site in a trigonal crystal field. Such an arrangement naturally describes, for example, the chalcogen layers of the 1T-TMDs (Fig.~\ref{f:overview}(a)). Fig.~\ref{f:overview}(c) summarises the splitting of the $p$-orbital energy levels as a result of bonding, crystal field splitting, and spin-orbit coupling. The bands that form from these will in general be anisotropic as the out-of-plane $p_z$ orbitals will have much larger hopping along the out-of-plane direction than the in-plane $p_{x/y}$ orbitals. For simplicity, we therefore initially neglect inter-layer hopping of the in-plane orbitals, leading to dispersionless $E$- ($p_{x/y}$)-derived levels as a function of the out-of-plane momentum, $k_z$. The $A_{1}$ ($p_z$-derived) bands, however, retain a strong $k_z$-dispersion (Fig.~\ref{f:overview}(d)). When the bandwidth arising due to inter-layer hopping becomes larger than the crystal field splitting (CFS), the $A_{1}$-derived band will cross through the $E$-derived ones, creating a set of $k_z$-dependent band inversions solely within the $p$-orbital derived manifold of states. In general, anti-crossing gaps can open at these intersections. This is indeed what should occur at the crossings of $R_{4}^\pm$ with $R_{4\prime}^\mp$ bands (Fig.~\ref{f:overview}(e)), as they both share the same symmetry character and angular momentum $m_J=1/2$. They have opposite parity, however, and thus their hybridization leads to an inverted band gap with a $\mathbb{Z}_2$ topological order. Accordingly, these gaps can be expected to host topological surface states, as we demonstrate below.
 
In contrast, the $R_{4}^\pm$ and $R_{5,6}^\mp$-derived bands belong to different irreducible representations. As a result, they behave differently under the application of the rotational operator $C_{3v}$ (see Supplementary Fig.~S1), and their crossing is protected against hybridization as long as it occurs at a $k$-point with $C_{3v}$ symmetry and the host system has both inversion and time-reversal symmetries.\cite{young_dirac_2012,yang_classification_2014, yang_prb_2015}   For the model considered here, this is satisfied for all $k$-points along the $\Gamma$-A direction of the three-dimensional Brillouin zone ($k_x=k_y=0$, varying $k_z$, see Fig.~\ref{f:overview}(b)). Consequently, the crossing of the $R_{4}^\pm$ and $R_{5,6}^\mp$-derived bands will lead to a single point of degeneracy (i.e., a bulk Dirac point) located part-way along this direction. Its location in momentum space is set both by the bandwidth of the $R_{4}^\pm$-derived band and by the strength of the CFS.

\begin{figure*}
\begin{center}
\includegraphics[width=\textwidth]{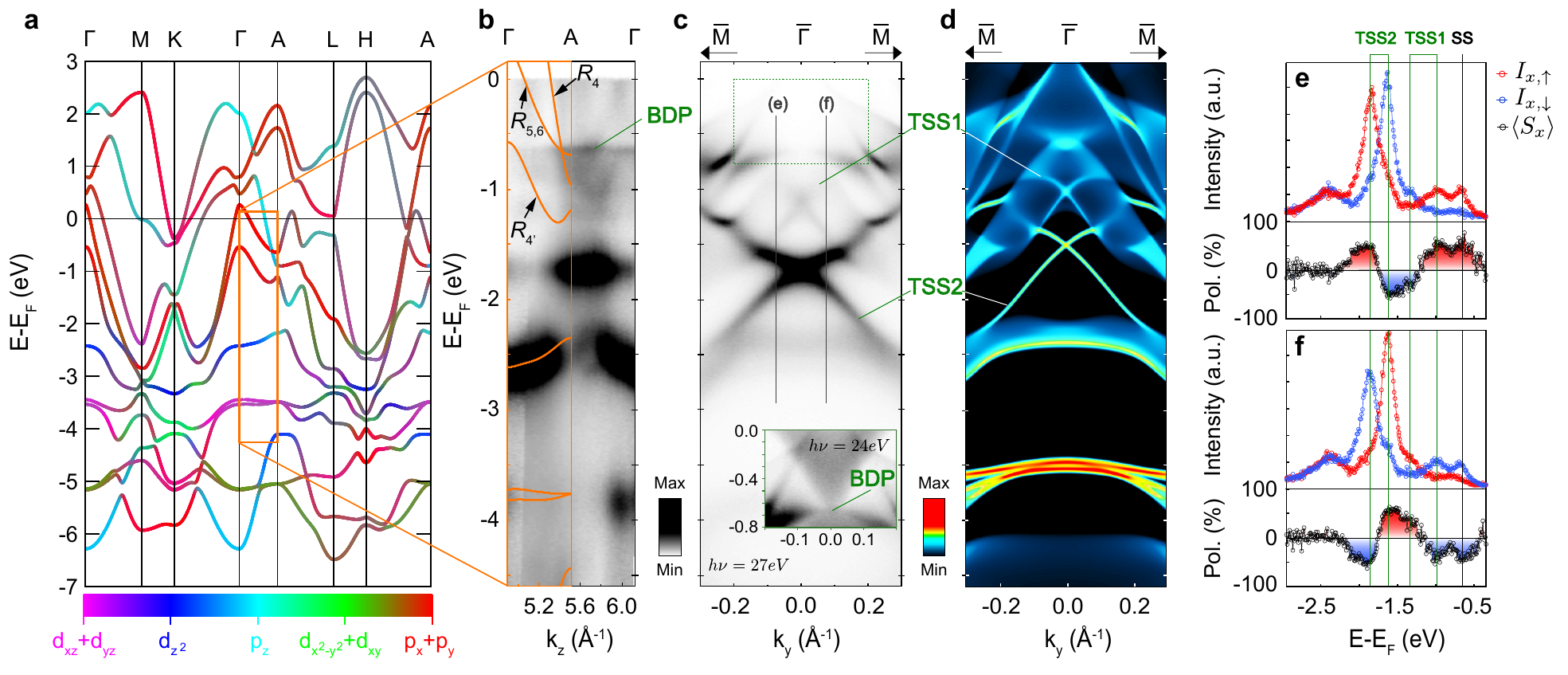}
\caption{ \label{f:PdTe2} {\bf Chalcogen-derived topological ladder in PdTe$_2$.} (a) Orbitally-resolved bulk electronic structure of PdTe$_2$, indicating dominantly chalcogen-derived orbital character for the states in the vicinity of the Fermi level. (b) Our ARPES measurements ($h\nu=80-132$~eV, $k_x=k_y=0$) reproduce the calculated out-of-plane dispersion when the calculations are rescaled by a factor of 1.08 (solid lines), revealing the formation of bulk Dirac points (BDPs) and gapped crossings of the upper $p_z$ and $p_{x,y}$-derived states. The corresponding bulk Dirac cones and topological surface states located within the inverted band gaps are clearly observed (c) in our ARPES measurements ($h\nu=$ 27~eV (24~eV for inset)) and (d) supercell calculations (projected onto the first 2 unit cells, see methods) along the $\overline{\Gamma}$-$\overline{\mbox{M}}$ direction. (e,f) Spin-resolved energy distribution curves along the lines shown in (c) reveal a clear helical spin texture of the two topological surface states (TSS1 \& 2), with an up-down-down-up relative ordering, as well as an additional spin-polarised state above TSS1 which we label SS.}
\end{center}
\end{figure*} 

In the schematic shown here (Fig.~\ref{f:overview}(e)), one branch of the Dirac cone is highly dispersive along $k_z$ while the other is completely dispersionless. This would place such Dirac cones exactly on the boundary of a maximally-tilted `conventional' (i.e. type-I) Dirac cone and an over-tilted one (i.e. a type-II bulk Dirac cone, in analogy to the recent classification of type-II Weyl fermions~\cite{soluyanov_type-ii_2015,Xu_structured}). In reality, the $R_{5,6}^\mp$-derived band will still have a finite, if small, out-of-plane dispersion. The group velocity of this band will determine whether a strongly titled type-I or type-II Dirac cone is obtained. 

\

\noindent{\bf Bulk Dirac points and topological surface states in PdTe$_2$}\\
We show in Fig.~\ref{f:PdTe2} that this simple model can be realised surprisingly well in the electronic structure of the TMD superconductor~\cite{raub_occurrence_1965} 1T-PdTe$_2$ (space group: $P\overline{3}m1$). The bands near the Fermi level are almost exclusively Te-derived (see also Supplementary Fig.~S1). Along $\Gamma$-$\mbox{A}$ (Fig.~\ref{f:PdTe2}(a)), two pairs of predominantly Te $p_{x,y}$ bands are evident within the energy region $E-E_F\sim\!-1$ to $\sim\!2$~eV (red colouring in Fig.~\ref{f:PdTe2}(a)), which we assign as the crystal-field and spin-orbit split bonding and anti-bonding $E$ bands in analogy with Fig.~\ref{f:overview}. They have modest out-of-plane dispersion, although much more significant dispersion can be observed along the in-plane directions consistent with their in-plane orbital character. In contrast, the $p_z$- ($A_{1}$)-derived states (cyan colouring in Fig.~\ref{f:PdTe2}(a)) have a dispersion along $\Gamma$-$\mbox{A}$ that spans nearly the entire valence band bandwidth, and thus crosses through the $E$ states as a function of $k_z$. 

Above the Fermi level, where the $R_{4}^-$ band intersects the anti-bonding $R_{5,6}^+$ and $R_{4\prime}^+$ states, a clear type-I protected crossing (upper) and an avoided crossing (lower) are formed, respectively. A similar phenomenology is observed for the bands immediately below $E_F$: the first crossing of the $p_z$-derived band with the bonding $R_{5,6}^-$ states leads to another protected BDP, this time of type-II character (see also Supplementary Fig~S2). The second crossing is again gapped. In fact, the proximity of this final crossing to both the anti-bonding and bonding-like branches of the $p_z$-derived bands causes an additional inverted gap to open directly below this. The deeper one ($E-E_F\sim\!-1.7$~eV in Fig.~\ref{f:PdTe2}(a,b)) is generated directly by the anti-crossing of bonding $R_4^+$ and $R_4'^-$ states, evident from a small kink structure near the A-point of the $R_4'$ band. The shallower band gap ($E-E_F\sim\!-1.1$~eV in Fig.~\ref{f:PdTe2}(a,b)) results from the crossing of bonding $R_4'$ with both anti-bonding $R_4$ and bonding $R_4$ states. As the latter two states have opposite parities the total parity of the lower band at the A-point becomes opposite to that at the $\Gamma$-point (see Supplementary Fig.~S1 for an explicit calculation of band parities), and hence this is also an inverted band gap with $\mathbb{Z}_2$ topological order.

\begin{figure*}
\begin{center}
\includegraphics[width=\textwidth]{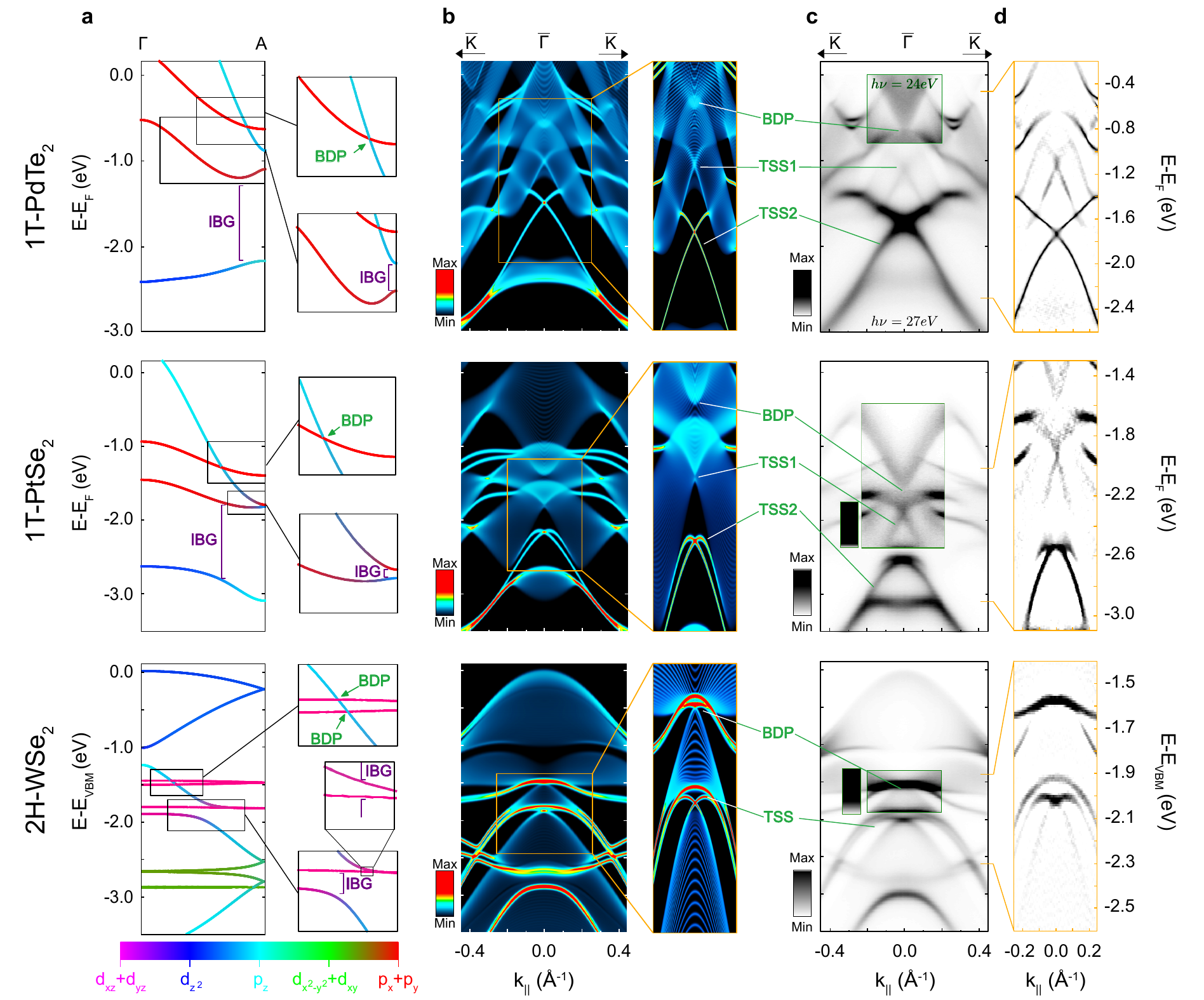}
\caption{ \label{f:TMDs} {\bf Generic observation of bulk Dirac fermions and topological surface states in TMDs. } (a) Orbitally-resolved out-of-plane bulk electronic structure of (top to bottom) PdTe$_2$, PtSe$_2$, and WSe$_2$, revealing the formation of bulk Dirac points (BDPs) and inverted band gaps (IBGs) as discussed in the text. (b) Surface-projected supercell calculations (along $\overline{\Gamma}-\overline{\mbox{K}}$), (c) ARPES measurements (top to bottom: $h\nu=27$~eV, $p$-pol; $h\nu=64$~eV, $p$-pol, $h\nu=49$~eV, CR+CL polarisation) and (d) corresponding curvature analysis\cite{zhang_precise_2011} show the surface-projected electronic structure of each compound, revealing the presence of the bulk Dirac cones as well as topological surface states located within the IBGs. The insets in (c) show the ARPES data measured with a different photon energy (PdTe$_2$, $h\nu=24$~eV) or shown with a different colour contrast (PtSe$_2$ and WSe$_2$) to better highlight some key features of the data. }
\end{center}
\end{figure*}

These features are well reproduced by our photon energy-dependent angle-resolved-photoemission (ARPES) measurements of the occupied electronic structure (Fig.~\ref{f:PdTe2}(b)). While the measured spectral features are broadened due to the finite $k_z$-resolution of photoemission, a significant $k_z$ dispersion of a number of states along $\Gamma$-A can still be observed. In the vicinity of $E_F$, we observe a light and more massive band which cross leading to an enhanced spectral weight at a binding energy of $\sim\!0.65$~eV close to the bulk A-point along $k_z$. The in-plane dispersion of these same states (insets of Fig.~\ref{f:PdTe2}(c) and Fig.~\ref{f:TMDs}(c) and Supplementary Fig.~S3) reveal diffuse ``filled-in'' intensity (again due to finite $k_z$-resolution) forming the upper part of this Dirac cone. Together, these observations and calculations therefore firmly identify the presence of type-II Dirac cones in PdTe$_2$,~\cite{pdte2footnote} arising due to the protected crossing of Te $p_z$- and $p_{x,y}$-crystal field-split states as they disperse differently with out-of-plane momentum. We note that spectroscopic signatures of the bulk Dirac cone extend up to the Fermi level and hence these Dirac fermions may carry signatures in transport measurements.~\cite{fei_nontrivial_2016}  

Additional states which are non-dispersive in $k_z$, and thus two-dimensional, are also evident in Fig.~\ref{f:PdTe2}(b). Most prominent is a band visible at $E-E_F\sim\!-1.7$~eV, an energy at which no bulk states are present along $\Gamma$-A. We thus assign this as a surface state. Its in-plane dispersion (Fig.~\ref{f:PdTe2}(c) and Supplementary Fig.~S4) shows a clear Dirac-like dispersion in the vicinity of $\overline{\Gamma}$, and is well reproduced by our supercell calculations of the surface electronic structure (Fig.~\ref{f:PdTe2}(d) and Supplementary Fig.~S5, see Methods), confirming its surface-derived origin. This has recently been observed by Yan~{\it et al.}~\cite{yan_identification_2015} and assigned as a topological surface state. Our measurements and calculations fully support this assignment: we find that it is located within the $k_z$-projected band gap that arises from the lower of the two avoided crossings below the Fermi level, between the $R_{4}^+$ and $R_{4\prime}^-$ bands identified above. To definitively identify its topological nature, we perform additional spin-resolved ARPES measurements (Fig.~\ref{f:PdTe2}(e) and Supplementary Fig.~S6). These reveal that this state is strongly spin-polarised (from fits to energy distribution curves (EDCs), we find an in-plane spin polarisation of $92\pm{14}$\% ($73\pm{16}$\%) for the upper (lower) branch of this surface state). The spin lies almost entirely within the surface plane and is locked perpendicular to the in-plane momentum, thus exhibiting the helical spin texture that is a defining characteristic of surface states of topological insulators, as also found from our supercell calculations (Supplementary Fig.~S4(c)). We refer below to this topological surface state as TSS2.

More subtly, our supercell calculations also reveal an additional surface-localised state forming another two-dimensional Dirac cone-like feature located at the energy of the band gap opened by the crossing of the $R_{4}^-$ and $R_{4\prime}^-$ bands. Unlike for TSS2, however, the band gap in the bulk spectrum opened by this avoided crossing does not span the entire Brillouin zone in $k_z$. The spectral weight of the surface-derived feature therefore lies within the manifold of projected bulk states which disperse around this avoided crossing. It is therefore better defined as a surface resonance rather than a true surface state. Consistent with this, we find that its wavefunction is more extended below the surface than for TSS2 (Supplementary Fig.~S5). Nonetheless, clear signatures of its in-plane Dirac-like dispersion are visible in our ARPES measurements at selected photon energies (Fig.~\ref{f:PdTe2}(c)), while our spin-resolved measurements (Fig.~\ref{f:PdTe2}(e)) reveal that it retains the spin-momentum locking characteristic of a TSS. Excitingly, therefore, our findings reveal how the band inversion created by the crossing of $p$-orbital $E$ and $A_1$-like bands in PdTe$_2$ drives the formation of a topological state (we refer to this as TSS1) whose topological origin still requires its existence despite the additional presence of bulk states at the same energies and in-plane momenta, thereby creating a topological surface resonance. 

Intriguingly, we find an additional two-dimensional state evident as a non-dispersive feature in Fig.~\ref{f:PdTe2}(b) that is pinned at exactly the energy of the bulk Dirac point. Tracking this state slightly away from the Dirac point along the $\overline{\Gamma}-\overline{\mbox{M}}$ in-plane direction, we find that it hosts a strong in-plane spin polarisation with the same sign as the upper branch of TSS1 (labeled SS in Fig.~\ref{f:PdTe2}(e,f); see also Supplementary Fig.~S6 which shows that this develops some out-of-plane spin canting along $\overline{\Gamma}-\overline{\mbox{K}}$). Spin-polarised Fermi arc surface states intersecting the Dirac point would naturally be expected for, e.g., the $(100)$ surface, where the bulk Dirac points project to different surface momenta (see Supplementary Fig.~S7).~\cite{xu_observation_2015,yi_evidence_2014} For the experimental $(001)$ cleavage plane, however, the two bulk Dirac points project exactly on top of each other and so such surface Fermi arcs would not naively be expected. Nonetheless, we note that topological surface states pinned to the Dirac point have recently been reported in calculations for other type-II bulk Dirac systems.~\cite{chang_typeii_2016} The origin of the states observed here therefore requires further investigation. Irrespective, the experimental observation of an additional spin-polarised surface state here stands as a further example of the rich surface electronic structure that this compound possesses, driven by an intricate array of band inversions within the $p$-orbital manifold of its bulk electronic structure.

\

\noindent{\bf Ubiquitous formation of BDPs and TSSs}\\
We show in Fig.~\ref{f:TMDs} and Supplementary Fig.~{S8} how such band inversions can be found in multiple other TMDs with different local and global crystalline symmetries, and which exhibit widely varying bulk properties. We first consider the closely-related compound, 1T-PtSe$_2$. This is semi-metallic, with a smaller overlap of chalcogen-derived bonding and anti-bonding states than in PdTe$_2$.~\cite{guo_electronic_1986} The transition metal states again contribute relatively little near to the Fermi level, while the $p_z$-derived chalcogen band can be clearly resolved cutting through the $p_{x,y}$-derived states in the vicinity of $E_F$ (Fig.~\ref{f:TMDs}(a)). A single type-II bulk Dirac cone and a pair of TSSs are stabilised in the occupied electronic structure just as for PdTe$_2$. These are evident in our supercell calculations (Fig.~\ref{f:TMDs}(b)) and well matched by our experimental ARPES measurements (Fig.~\ref{f:TMDs}(c,d)  and Supplementary Fig.~S9). The spin-orbit coupling of the Se manifold is weaker than that of Te, evident from both the smaller splitting between $E$-like states and from smaller anti-crossing gaps which open in the vicinity of unprotected band crossings. The local band gaps in which the TSSs reside are therefore smaller than in PdTe$_2$, causing the upper branches of the TSSs to rapidly ``turn over'' to maintain the surface-bulk connectivity as required by their topological origin. 

Nonetheless, in contrast to the common picture for well-known topological insulators such as Bi$_2$Se$_3$, the band inversions leading to such topological surface states, as well as the bulk Dirac cone formation, naturally survive this reduction in spin-orbit coupling strength. Indeed, the relevant energy scales for stabilising the topological surface states here are the $p_z$-derived bandwidth vs.\ the trigonal crystal field splitting. While increased spin-orbit coupling strength will open larger hybridisation gaps, our findings (see Fig.~\ref{f:TMDs}(c,d)) demonstrate how the topological surface states survive as topological surface resonances even in the limit where the hybridisation gap opened is significantly smaller than the dispersion of bulk electronic states around this. They should therefore be a very robust feature of the intrinsic $p$-orbital band inversions found here. The recent observation of a type-II BDP in PtTe$_2$~\cite{yan_lorentz-violating_2016} can also be understood within the same classification that we present here, establishing our findings as generic to the group-10 TMD metals and semi-metals.~\cite{huang_type-ii_2016} We further show in Supplementary Fig.~{S7} and Supplementary Fig.~{S8(a,b)} how such bulk band crossings/inversions also occur for the high-temperature 1T phase of the group-9 TMD IrTe$_2$. In this system, the trigonal symmetry which protects the BDP is lost upon cooling through a structural phase transition,\cite{cao_origin_2013,fang_structural_2013} raising prospects to investigate temperature-driven topological phase transitions and mass gap opening of the type-II Dirac fermions.  

Fig.~\ref{f:TMDs} shows how similar states are also stabilised for a different TMD polymorph: the 2H structure of WSe$_2$ (space group: $P63/mmc$). Our bulk band structure calculations along $k_z$ (Fig.~\ref{f:TMDs}(a)), which are in good agreement with previous photon energy-dependent ARPES measurements,~\cite{riley_direct_2014} reveal a strongly dispersive band with significant $p_z$ orbital character. This is intersected by very weakly dispersing bands at around 1.5 and 1.9~eV (2.7 and 2.9~eV) below the valence band top which we attribute as the anti-bonding (bonding) $E$-like bands, respectively. Unlike for PdTe$_2$, the Fermi level lies in a band gap of both the transition-metal (formally in a $d^2$ configuration) and chalcogen-derived states, and so this system is a semiconductor.\cite{chhowalla_2013,xu_spin_2014,riley_negative_2015} Moreover, transition-metal and chalcogen-derived states are no longer well separated in energy, and so the $E$-like bands have a strong transition-metal $d$-orbital character intermixed with their Se $p_{x,y}$ character. The more localised nature of the $d$ vs.\ $p$ orbitals, together with an increased inter-layer separation, leads to a significantly smaller out-of-plane dispersion of these $E$-like bands than for PdTe$_2$. Finally, the unit cell contains two MX$_2$ (M$=$transition metal, X$=$chalcogen) layers in the 2H structure, as compared to a single such layer in the 1T structure. This results in an effective backfolding of the bands about the Brillouin zone boundary along $k_z$, doubling each of the $R_{5,6}^\pm$ and $R_{4\prime}^\pm$ bands as seen in our calculations. 

The C$_{3v}$-symmetry enforced degeneracy of the $R_{4}$-$R_{5,6}$ crossings discussed above, however, still holds. Now, therefore, the crossing of the dispersive $R_4$ band with each of the back-folded $R_{5,6}$ bands leads to a pair of closely-spaced bulk Dirac cones. These are almost maximally tilted and, unlike for PdTe$_2$,  now additionally host significant transition-metal character at the BDP. Intriguingly, as the back-folding by definition changes the sign of the band's group velocity, this leads to stacked Dirac points of opposite character (type-II and type-I for the upper and lower crossings, respectively). We observe clear spectral signatures of the in-plane dispersion of these Dirac cones (Fig.~\ref{f:TMDs}(c)), but are unable to resolve a splitting of the two cones experimentally due to their small energy separation and strong three-dimensional dispersions. Both crossings of the $R_{4}$ and back-folded $R_{4\prime}$ bands become gapped, and would therefore be expected to host topological surface states/resonances as in PdTe$_2$. One such band gap is too small to resolve experimentally, while for the lower branch a clear inverted band gap is opened. Our supercell calculations (Fig.~\ref{f:TMDs}(b)) indeed reveal the TSS located within this band gap, spanning between the manifold of bulk states above and below the band gap. Although the resulting band gap is small, the in-plane dispersion is significant. Our ARPES and spin-ARPES measurements (Fig.~\ref{f:TMDs}(c) and Supplementary Fig.~S10) show clear evidence for the existence of the resulting surface state, its band-gap crossing nature, and its chiral spin polarisation. As shown in Supplementary Fig.~S8(c-f), we find similar bulk Dirac cones and inverted band gaps in other 2H-structured TMDs, TaSe$_2$ and NbSe$_2$ (space group: P63/mmc), despite them hosting a different layer stacking sequence as compared to WSe$_2$. This opens the exciting prospect to investigate the influence of charge order, which these compounds host~\cite{wilson_charge-density_1974,borisenko_pseudogap_2008,yokoya_fermi_2001,bawden_spin-valley_2016}, and the consequent reconstruction of the electronic structure, on the topological and bulk Dirac states. 

\begin{figure}[!h]
\includegraphics[width=\columnwidth]{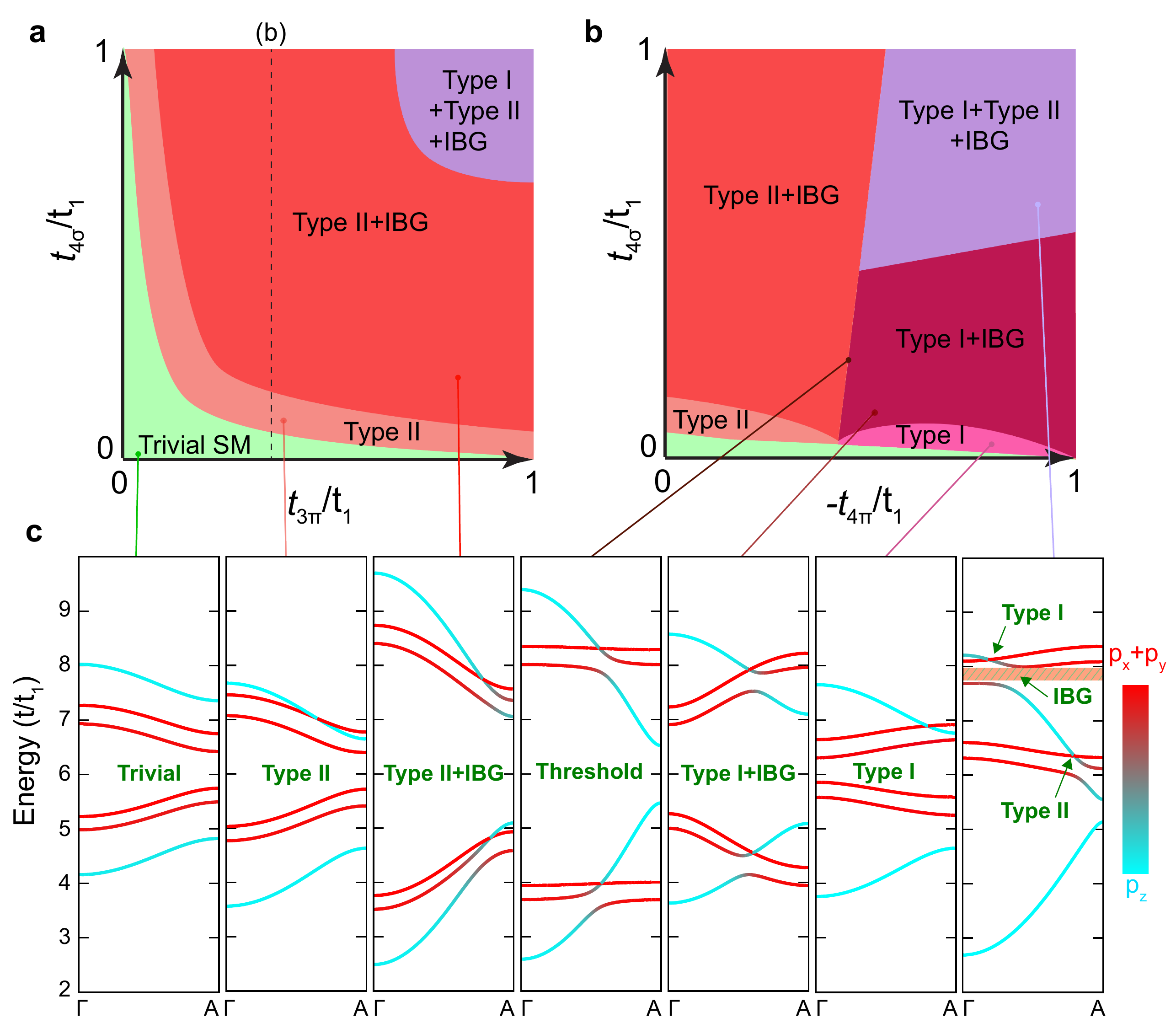}
\caption{ \label{f:phase} {\bf Interlayer hopping-controlled topological and Dirac phases.} (a,b) Effective phase diagrams for a minimal 2-site $p$-orbital tight-binding model (see Methods), indicating the tuneability of Dirac crossings and band inversions as a function of intra-layer hopping ($t_1=t_2$), inter-layer hopping within the unit cell ($t_3$), and inter-layer hopping between unit cells ($t_4$). As the inter-layer hopping is increased, the band width of the $p_z$-derived band grows rapidly such that it overcomes the crystal field splitting and spin-orbit split-off energies of the $p_{x/y}$-derived states. (a) Considering only $\sigma$-type bonding between unit cells, this drives successive transitions from a trivial semimetal to a type-II Dirac state to a system supporting a type-II Dirac fermion and an inverted band gap. (c) This is shown explicitly in example electronic structure calculations along $\Gamma$-A (left three panels, calculated for the points indicated in (a)). An additional type I Dirac cone is found for large inter-layer hopping. (b) A richer phase diagram is obtained when including non-zero $\pi$-type interactions for inter-unit cell hopping, shown as a function of increasing $t_{4\pi}$ with $t_{3\pi}$ and $t_{4\sigma}$ taking values along the black dashed line indicated in (a). Such a $\pi$-type hopping enhances the interaction between the $p_{x,y}$ orbitals of neighboring cells and accordingly can influence and even reverse the slope of the corresponding $p_{x,y}$ bands. Multiple band inversions and Dirac points of both type-I and type-II character, as well as their combination as found in e.g. PdTe$_2$, are obtained (right four panels in (c)).} 
\end{figure}

\

\noindent{\bf Tuneability and robustness against inversion symmetry breaking}\\
The principle underlying the formation of bulk Dirac cones and topological surface states here is very general, and can be expected to occur across numerous materials systems. Moreover, our demonstration of their existence across multiple TMDs indicates that there is still significant opportunity to tailor the properties, locations, and nature of these states. To show this explicitly, we construct a tight-binding model for our minimal 2-site system considered in Fig.~\ref{f:overview}. Fig.~\ref{f:phase} shows how varying the inter-layer hopping both within and between neighbouring unit cells, as well as adjusting the ratio of $\sigma$-type and $\pi$-type inter-layer interactions, leads to a rich array of coexisting topological states and phases. Controlling these experimentally should be possible by varying the degree of covalency in the system and tuning the out-of-plane lattice parameter via atomic substitution or applied uniaxial pressure or strain along the $c$-axis. Such a strain field would not affect the trigonal symmetry which protects the Dirac points within the inverted phases, but could be used to traverse the phase boundaries, providing powerful routes to tuneable topological phase transitions and the creation or annihilation of bulk Dirac points in TMDs. 

Moreover, the insights gained here suggest strategies for the design of Dirac and topological phases. As an illustration of this, we consider replacing one of the Te layers in PdTe$_2$ by Se.  In contrast to PdTe$_2$, this structure is non-centrosymmetric.  Typically, such a loss of inversion symmetry would be assumed to lift the spin degeneracy, splitting the Dirac point into a pair of Weyl points.  In contrast, since the PdTeSe structure we consider retains trigonal symmetry, we find that both spin-degeneracy and the protected Dirac crossing are maintained along the rotational axis ($\Gamma$-A), but spin degeneracy is lost elsewhere (Supplementary Fig.~S11). The Dirac point in this case can therefore be considered as a protected degeneracy of two Weyl points that would not typically be expected. Our study thus opens routes to the rational design of topological materials, and indicates just how wide a purview topological band structure effects can be expected to have.
\

\noindent{\bf Acknowledgements}
We thank R.~Arita and N. Nagaosa for useful discussions and feedback and F. Bertran and P. Le F\`evre for ongoing technical support
of the CASIOPEE beam line at SOLEIL. We gratefully acknowledge support from the CREST, JST (Nos. JPMJCR16F1-16F2), the Leverhulme Trust, the Engineering and Physical Sciences Research Council, UK (Grant Nos.~EP/M023427/1 and EP/I031014/1), the Royal Society, the Japan Society for Promotion of Science (Grant-in-Aid for Scientific Research (S); No. 24224009 and (B); No. 16H03847), the International Max-Planck Partnership for Measurement and Observation at the Quantum Limit, Thailand Research Fund and Suranaree University of Technology (Grant No. BRG5880010) and the Research Council of Norway through its Centres of Excellence funding scheme, project number 262633, “QuSpin”, and through the Fripro program, project number 250985 “FunTopoMat”. This work has been partly performed in the framework of the nanoscience foundry and fine analysis (NFFA-MIUR Italy, Progetti Internazionali) facility. B.-J. Y. was supported by the Institute for Basic Science in Korea (Grant No. IBS-R009-D1), Research Resettlement Fund for the new faculty of Seoul National University, and Basic Science Research Program through the National Research Foundation of Korea (NRF) funded by the Ministry of Education (Grant No. 0426-20150011). OJC, LB, JMR and VS acknowledge EPSRC for PhD studentship support through grant Nos.~EP/K503162/1, EP/G03673X/1, EP/L505079/1, and EP/L015110/1. IM acknowledges PhD studentship support from the IMPRS for the Chemistry and Physics of Quantum Materials. We thank Diamond Light Source (via Proposal Nos.~SI9500, SI12469, SI13438, and SI14927) Elettra,  SOLEIL, and Max-Lab synchrotrons for access to Beamlines I05, APE, CASSIOPEE, and i3, respectively, that contributed to the results presented here.

\noindent{\bf Author Contributions.} MSB and BJY performed the theoretical calculations. The experimental data was measured by OJC, JFe, LB, JMR, IM, FM, VS, DB, SPC, MJ, JWW, TE, WM, and PDCK, and analysed by OJC. ML, TB, JFu, IV, JR, TKK, and MH maintained the ARPES/SARPES end stations and provided experimental support. KO, MA, and TS synthesised the measured samples. PDCK, OJC, and MSB wrote the manuscript with input and discussion from co-authors. PDCK and MSB were responsible for overall project planning and direction.

\noindent{\bf Author Information.} Reprints and permissions information is available at www.nature.com/reprints. The authors declare no competing financial interests. Correspondence and requests for materials should be addressed to PDCK or MSB.

\
\clearpage
{\small 
\noindent{\bf Methods}\\

\noindent{\bf Calculations:} 
 The bulk calculations were performed within density functional theory (DFT) using Perdew-Burke-Ernzerhof exchange-correlation functional as implemented in the WIEN2K program.~\cite{wien2k} Relativistic effects including spin-orbit coupling were fully taken into account. For all atoms, the muffin-tin radius $R_{MT}$ was chosen such that its product with the maximum modulus of reciprocal vectors $K_{max}$ become $R_{MT} K_{max}=7.0$. The Brillouin zone sampling of 1T (2H) structures was carried out using a $20\times 20\times 20$ ($20\times 20\times 10$) $k$-mesh. 
For  the surface calculations, a 100 unit tight binding supercell was  constructed using maximally localized Wannier functions.~\cite{souza, mostofi, kunes}  The $p$-orbitals of the the chalcogen and the $d$-orbitals of the transition metal atoms were chosen as the projection centres.  

The phase diagrams and related band structures shown in Fig.~\ref{f:phase} were constructed using a 12-band tight-binding model, considering nearest-neighbour $p-p$ hoppings between the chalcogen sites in a trigonal unit cell similar to that of 1T-TMDs, but without any transition metal element. The basis set is accordingly composed of two sites, $j=1$ and $2$, and each site contains six spin-orbital components,  $|p_{i,j},\sigma\rangle$, where $i=x,y,z$ and $\sigma=\uparrow, \downarrow$. The hopping integrals $t_{ij,i\prime j\prime}=\langle p_{ij}|H|p_{i\prime j\prime}\rangle$ were calculated using the Salter-Koster method by choosing the appropriate values for on-site crystal field terms $\Delta_{CFS}$ and the two-centre bond integrals $t_{ii\prime\sigma}$ and $t_{ii\prime\pi}$.~\cite{Salter-Koster} For simplicity, the effect of spin-orbit interaction was approximated by only considering the on-site contribution $H_{so}=\lambda \bm{L}\cdot \bm{S}$, where $\bm{L}$ and $\bm{S}$ are orbital and spin angular momentum operators, respectively. Considering the hopping paths shown in Fig~\ref{f:overview}(a), each band structure calculation required setting eight hopping parameters 
$t_{k\sigma}$, $t_{k\pi}$ where $k=1-4$ as well as $\Delta_{CFS}$ and $\lambda$. We fix $t_{1\sigma}= t_{1\pi}=t_{2\sigma}= t_{2\pi}=1.0$, the crystal-field splitting, $\Delta_{CFS}=1$, and the spin-orbit coupling $\lambda=0.3$. Intra-unit inter-layer hopping is assumed to be of $\pi$-type only ($t_{3\pi}$ [$t_{3\sigma}=0$]). The other parameter were varied to produce the representative band structures shown in Fig.~\ref{f:phase}(c). Inter-unit cell hopping is assumed to be dominated by $p_z$ orbitals and is therefore predominantly of $\sigma$-type ($t_{4\sigma}$), although we also consider the effect of finite $\pi$-type interactions between neighbouring unit cells ($t_{4\pi}\ll{t_{4\sigma}}$).

\

\noindent{\bf ARPES:} ARPES measurements of PdTe$_2$ and PtSe$_2$ were performed at the I05 beamline of Diamond Light Source, UK, and most spin-integrated WSe$_2$ measurements at the CASSIOPEE beamline of Synchrotron SOLEIL, France. Additional ARPES measurements of WSe$_2$ were taken at the APE beamline of Elettra Syncrotrone Trieste, Italy, along with the majority of the spin-resolved ARPES measurements. Additional spin-resolved measurements of PdTe$_2$ were obtained from the I3 beamline of MAX IV Laboratory, Sweden.

High-quality single crystal samples, grown by chemical vapour transport, were cleaved \textit{in situ} at temperatures ranging between 9-15K. Measurements were performed using either p-polarised (PdTe$_2$, PtSe$_2$, WSe$_2$), or circularly polarised (WSe$_2$) light, and using photon energies in the range $h\nu=24-132$~eV. Scienta R4000 hemispherical analysers, with a vertical entrance slit and the light incident in the horizontal plane, were used at Diamond and SOLEIL. 

A VG-Scienta DA30 analyser (Elettra), fitted with two very low energy electron diffraction (VLEED) based spin polarimiters~\cite{bigi_very_2017}, was utilised for the majority of the spin-resolved measurements along three momentum directions, while additional measurements were performed using a mini-Mott setup on a Scienta R4000 analyser (Max IV). The finite spin-detection efficiency was corrected using detector-dependent Sherman functions ranging between $S = 0.17 \pm 0.03$ and $S = 0.43 \pm 0.03$ as determined by fitting the spin-polarisation of reference measurements of the Au$(111)$ Rashba-split surface state for each experimental set-up utilised. Spin-resolved EDCs were determined according to 

\begin{equation}
	I_i^{\uparrow,\downarrow}=\frac{I_i^{tot} (1\pm P_i )}{2},
\end{equation}
where
\begin{math}
	i=\{x,y,z\}, \quad
	I_i^{tot}=(I_i^++I_i^-)
\end{math} and
$I_i^\pm$ is the measured intensity for a positively or negatively magnetised detector, corrected by a relative efficiency calibration. The final spin polarisation is defined as follows:

\begin{equation}
P_i=\frac{I_i^+-I_i^-}{S(I_i^++I_i^-)},
\end{equation}
where S is the relevant Sherman function for the detector in use.

Quantitative spin-polarisation magnitudes were determined from the relative areas of Lorentzian peak fits to energy distribution curves (EDCs) originating from oppositely magnetised detectors. A Shirley background and Gaussian broadening were included in this analysis.

To determine the PdTe$_2$ $k_z$ dispersion from photon-energy-dependent ARPES, we employed a free electron final state model
\begin{equation}
k_z=\sqrt{\frac{2m_e}{\hbar^2}} (V_0+E_k  \cos^2\theta )^{1/2}
\end{equation}
where $\theta$ is the in-plane emission angle and $V_0$ is the inner potential. We find best agreement to density-functional theory calculations taking an inner potential of 16 eV and a $c$-axis lattice constant of 5.13 \AA.

\ 

\noindent{\bf Data availability statement:} The data that underpins the findings of this study are available at http://dx.doi.org/10.17630/27a2dc90-470f-4e69-be1e-5ebb072db739.

\

\end{document}